\documentclass[manuscript,natbib209]{aastex}

\usepackage{footnote}
\usepackage{float}
\usepackage{multirow}

\begin{document}

\title{Optical and Radio Observations of the T Tauri Binary KH 15D (V582 Mon): Stellar Properties, Disk Mass Limit and Discovery of a CO Outflow}


\author{Rachel A. Aronow, William Herbst, and A. Meredith Hughes}
\affil{Department of Astronomy, Van Vleck Observatory, Wesleyan University, 96 Foss Hill Drive
    Middletown, CT 06459, USA}

\author{David J. Wilner}
\affil{Harvard-Smithsonian Center for Astrophysics, 60 Garden Street, 
       Cambridge, MA 02138, USA}
    
\and

\author{Joshua N. Winn}
\affil{Department of Astrophysical Sciences, Princeton University, 4 Ivy Lane, Princeton, NJ 08540, USA}



\begin{abstract}
    We present VRIJHK photometry of the KH 15D T Tauri binary system for the 2015/16 and 2016/17 observing seasons. For the first time in the modern (CCD) era we are seeing Star B fully emerge from behind the trailing edge of the precessing circumbinary ring during each apastron passage. We are, therefore, able to measure its luminosity and color. Decades of photometry on the system now allow us to infer the effective temperature, radius, mass and age of each binary component. We find our values to be in good agreement with previous studies, including archival photographic photometry from the era when both stars were fully visible, and they set the stage for a full model of the system that can be constructed once radial velocity measurements are available. We also present the first high sensitivity radio observations of the system, taken with ALMA and the SMA. The respective 2.0 and 0.88 mm observations provide an upper limit on the circumbinary (gas and dust) disk mass of 1.7 M$_\mathrm{Jup}$ and reveal an extended CO outflow which overlaps with the position, systemic velocity and orientation of the KH 15D system and is certainly associated with it. The low velocity, tight collimation, and extended nature of the emission suggest that the outflow is inclined nearly orthogonal to the line of sight, implying it is also orthogonal to the circumbinary ring. The position angle of the radio outflow also agrees precisely with the direction of polarization of the optical emission during the faint phase. A small offset between the optical image of the binary and the central line of the CO outflow remains as a puzzle and possible clue to the jet launching mechanism. 
\end{abstract}




\section{Introduction}
KH 15D (V582 Mon) is a binary T Tauri member of the young cluster NGC 2264 that exhibits unique optical variability discovered by \citet{KearnsHerbst} at Wesleyan University's Van Vleck Observatory. Extensive monitoring during the late 1990's revealed that its brightness changed by nearly 3.5 magnitudes and that the variation was eclipse-like, strictly periodic with P = 48.37 days, but included a central brightness reversal and an eclipse duration that was lengthening with time \citep{Hamilton2001}. It was clear early on from the duration, structure and evolution of the eclipses in this system that they must be caused by circumstellar matter, not by a star \citep{Herbst2002}.  \citet{ChiangMurrayClay} and \citet{Winn2004} independently proposed a binary model for the object, in which only one of the stellar components was directly visible, while the companion's orbit was entirely obscured by a circumbinary (CB) ring. The periodically visible component was denoted Star A, and its unseen companion was called Star B. This model was confirmed by radial velocity measurements of Star A \citep{Johnson2004}, which showed that it was indeed a spectroscopic binary. As the stars traveled along their eccentric orbits ($e\approx0.6$), Star A was `rising' above and `setting' behind the highly opaque and sharp-edged ring during each orbital cycle, while Star B remained occulted by the ring at all times \citep{Winn2006}. Panel 2 of Figure \ref{figure:intro_figure} is an illustration of the geometry of the system at that time.

\begin{figure}[h!]
\centering
\epsscale{1}
\plotone{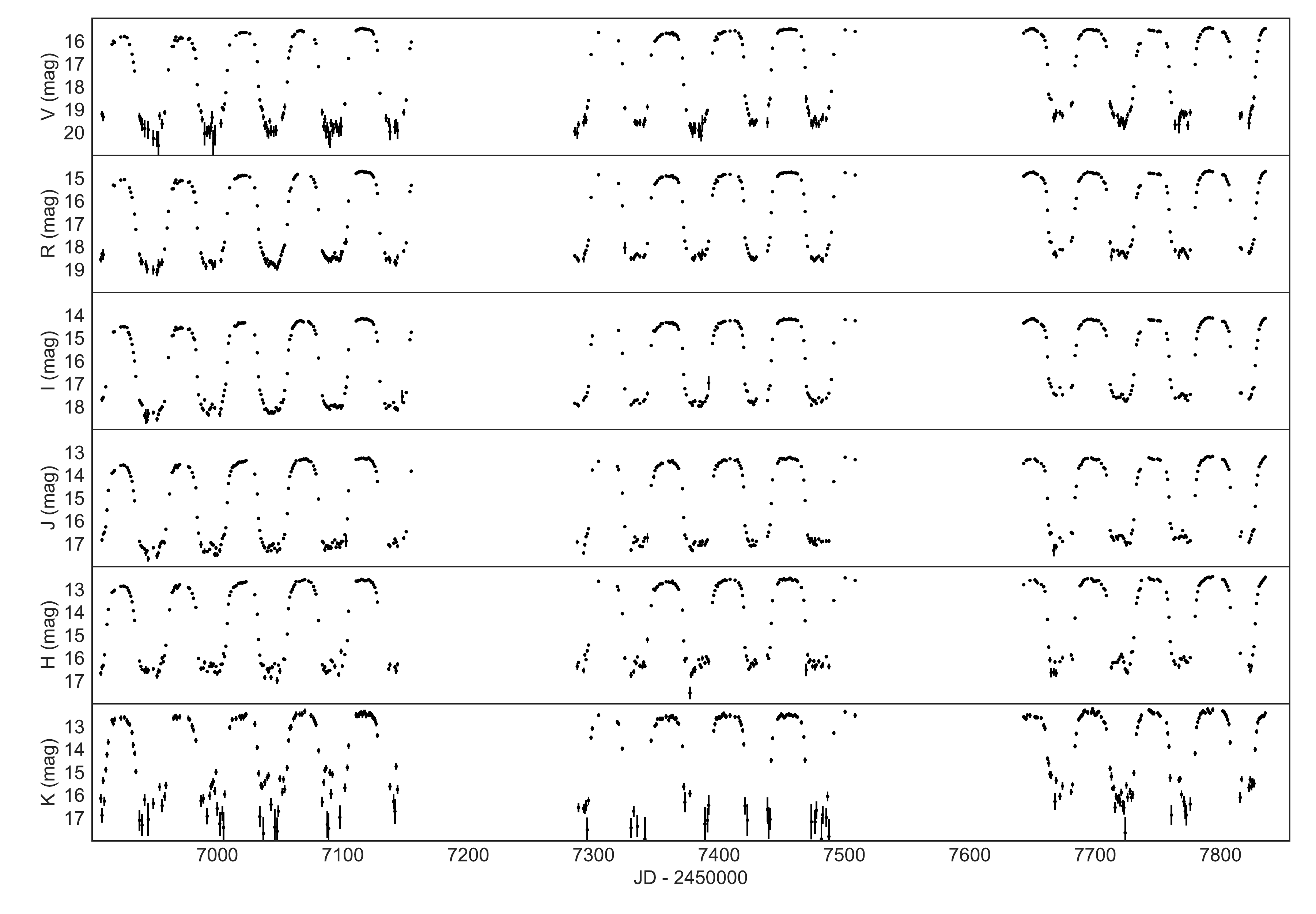}
\caption{ANDICAM photometry in VRIJHK of KH 15D, taken with a nightly cadence at CTIO over the last three observing seasons. After steadily increasing over the previous four years, the peaks of the light curve have now leveled off over the last one and a half years, signaling that the entirety of Star B is now completely unocculted at apastron.
\label{figure:ANDICAM_lightcurve}}
\end{figure}

In addition to the 48.37 day orbital cycle (see Figure \ref{figure:ANDICAM_lightcurve}), the system's brightness variations are modulated on a longer timescale. Throughout the 2000's, the periodicity of the eclipses were consistent, but the overall magnitude of the system began decreasing at all phases of the binary orbit. This long-term variation is now understood to be due to a rigid precession of the CB ring \citep{ChiangMurrayClay, Winn2004,Winn2006}. Viewed from Earth, the precession resulted in a sharp-edged opaque screen slowly moving across the plane of the sky, covering more and more of Star A's orbit with time, thereby increasing the duration of the eclipses. During the four-year period from 2006--2009, less and less of Star A's photosphere was visible above the horizon defined by the CB ring edge. By 2010, the entirety of Star A was occulted by the disk at all orbital phases (see panel 3 of Figure \ref{figure:intro_figure}), and the system's peak brightness was over 2 magnitudes dimmer than it had been in the late 1990's \citep{Herbst2010}. The system at this time was seen only by scattered light from one or both stars off of the ring edges. Nonetheless, the 48.37 day orbital period was still clearly visible in the photometry. 

While the \citet{ChiangMurrayClay} and \citet{Winn2004, Winn2006} models each accurately predicted the existence of this dim phase, neither predicted its duration successfully. A distinct brightening in the light curve was detected just two years after the dim phase began, suggesting that the previously unseen Star B had already started to peek out from behind the trailing edge of the disk \citep{Capelo2012}. By late 2016, the entirety of Star B was visible at each apastron passage, while Star A remained occulted at all times. We have entered a new phase in our view of this complex system, as illustrated in panel 4 of Figure \ref{figure:intro_figure}. 

The existence of the occulting CB ring and its steady precession makes KH 15D similar to a double-lined eclipsing binary in terms of the information on its stellar components that can be derived from it. Furthermore, it is one of only two such systems known to still be embedded in an accretion disk \citep{Gillen2017}. This makes it important to characterize as fully as possible not only the stellar properties of the system, but also the CB ring and accretion disk. The first attempt to directly detect dust continuum emission from the precessing CB ring or accretion disk at radio wavelengths was on December 15, 2001, using the Owens Valley Radio Observatory (OVRO). No point source was detected at this wavelength, placing a $\sim$1 mJy/beam upper limit on the 3 mm flux density of the ring or disk \citep{HamiltonThesis}. Studies of the CB ring at near-IR wavelengths are discussed in \citet{Nicole, Nicole2}. The system shows no detectable IR excess emission out to 8 $\mu$m, but inferences about the ring particle sizes and composition can be made from photometric and spectral variations as a function of orbital phase. The authors conclude that the data are consistent with scattering from 1-50 $\mu$m sized particles composed of water and methane ices. When the system is at its faintest in the optical there is some, as yet inconclusive, evidence in both the photometry and spectrscopy of a third source of emission -- possibly a luminous giant planet, brown dwarf or very low mass stellar companion \citep{WindemuthHerbst2014, Nicole2}.

In addition to the obscuring CB ring of solids responsible for its unique photometric characteristics, it is clear that KH 15D must still be embedded in a gaseous accretion disk. While it may be classified as a weak-lined T Tauri Star (WTTS) on the basis of a relatively small equivalent width of the H$\alpha$ line when Star A is out of eclipse \citep{Hamilton2001}, and the lack of a near-IR excess \citep{Nicole} at any orbital phase, \citet{Hamilton2003} presented unequivocable evidence that the system is still undergoing weak accretion. Spectra taken in both the eclipsed and out of eclipse phases show double-peaked forbidden lines of [OI]$\lambda$6300  and $\lambda$6363. The symmetry and low velocities of the forbidden line emission are indicative of a bipolar outflow inclined nearly perpendicular to the line of sight \citep{Hamilton2003, Mundt2010}. There are also broad and variable high velocity (up to 300 km/s) wings of the H$\alpha$ line that are indicative of magnetospheric accretion \citep{Hamilton2003, Hamilton2012}. While the rate may be 10 times less than for a typical classical TTS, there is no doubt that accretion continues.  

Further evidence for accretion driven outflow from the system comes from the presence of near-IR spectral features from shocked H$_2$ and CO \citep{Tokunaga2004, Deming2004,Nicole2}. \citet{Tokunaga2004} imaged KH 15D through a narrow band H$_2$ filter and detected a thin, well-collimated filamentary jet that appears to originate from the binary or close to it and extend to the north. At the time of its discovery, it was unclear if this linear structure was further evidence of the jet proposed by \citet{Hamilton2003}, or simply a circumstantial superposition of a molecular cloud filament. Particularly notable was the lack of a clearly visible counterjet to the south, although a faint hint of it was present. \citet{Deming2004} observed an extended double-peaked H$_2$ emission velocity profile centered on the location of KH 15D, providing more evidence of an association between KH 15D, the proposed outflow, and the H$_2$ structure.

In this paper, we present two more seasons of VRIJHK monitoring of the system, which brings us fully into the era in which Star B is unobscured at apastron passage. We combine the new data with all previously published optical and near-infrared photometry of KH 15D to determine the luminosities, radii, and effective temperatures of the binary components. We compare these results with the archival photometric record that goes back to the 1960's and records a time when Stars A and B were both visible \citep{JohnsonWinn2004,Johnson2005,Maffei2005}. The photometric record is used to estimate the age and masses of the binary components by comparing their derived luminosities and effective temperatures with models of pre-main sequence (PMS) stars. 

Additionally, we describe the first extensive radio studies of the system, done with the Atacama Large Millimeter/submillimeter Array (ALMA) and Submillimeter Array (SMA). From these observations we derive constraints on the accretion disk mass. We further report the discovery of a CO outflow in the radio data that is spatially coincident with the shocked molecular gas seen in the near-IR and clearly resolves the counterjet that extends to the south. Analysis of these data provides strong new evidence that the outflow is associated with the KH 15D system, although the morphology continues to raise questions about the details of how the jet is launched, how it has evolved with time and whether it arises from the binary itself or a putative third component to the KH 15D system.   

\section{Observations \& Data Reduction}
\subsection{Optical and Near-IR CCD Data: SMARTS}
For convenience, we have compiled in Table \ref{table:photometry} all of the high quality (CCD) photometry of KH 15D that has been obtained since 1995. The previously unpublished data, spanning the interval 2015 Sept to 2017 March were obtained with the ANDICAM instrument on the 1.3 m telescope at Cerro-Tololo Inter-American Observatory (CTIO), operated by the SMARTS consortium. KH 15D is observed nightly during each observing season in three optical filters (VRI) and three near-infrared filters (JHK). These data are obtained within an interval of about 30 minutes and combined to yield a single nightly magnitude in each color. The observing and reduction process is identical to what has been used for many years and is described in more detail by \citet{WindemuthHerbst2014}. The full data set is most complete in the I magnitude, which is plotted in Figure \ref{figure:intro_figure}. The newest data, for the last three observing seasons, is shown in Figure \ref{figure:ANDICAM_lightcurve} with enough time resolution that the individual orbital cycles may be seen.

\subsection{Radio Observations: SMA}
\label{sec:obs}
The Submillimeter Array (SMA) on Mauna Kea, Hawaii was used in a compact antenna configuration to observe
KH 15D, primarily to search for possible 870~$\mu$m dust continuum emission. Table~\ref{tab:obs} provides 
a summary of the observational parameters for each of two runs. The first run
consisted of a short (2 hour) filler track on 2010 October 13. The 
setup used 128 channels (0.8125 MHz spacing) in each of 24 chunks that 
together spanned $\pm(4-6)$~GHz from a local oscillator (LO) frequency of 342.05 GHz. 
The second run was a full (7.5 hour) track on 2011 October 25. 
This observation took advantage of expanded correlator capability
that processed $\pm(4-8)$~GHz from an LO frequency of 342.05 GHz,
which included the spectral line CO J=3--2 (345.796 GHz) in the upper side band.
The weather conditions were good for both of these observations, with 
stable atmospheric phase and low to modest opacity at 225~GHz, as measured 
by the tipping radiometer at the nearby Caltech Submillimeter Observatory.
The basic observing sequence consisted of 2.5 or 3 minutes on each of
the calibrators J0532+075 and J0750+125, and 10 minutes on KH 15D.  Passband 
calibration was performed using observations of available strong sources, 
and the absolute flux scale was set using observations of Callisto and Uranus, 
with an estimated accuracy of 20\%.  The field of view is set by the
$36''$ FWHM primary beam size at the observing frequency. The synthesized
beam sizes are $\sim2''$.  All of the calibration was performed 
using the \texttt{MIR} software package. Continuum and spectral line imaging 
and CLEAN deconvolution were done in the \texttt{miriad} package.

\subsection{Radio Observations: ALMA}
Atacama Large Millimeter/submillimeter Array (ALMA) Band 4 observations of KH 15D were obtained on January 16, 2016 (\dataset{ADS/JAO.ALMA\#2015.1.01537.S.}). The object was observed for a total of 11.2 minutes using forty-six 12-m antennas. The data set contains four spectral windows: spw0 (138.99 -- 137.00 GHz), spw1 (140.92 -- 138.94 GHz), spw2 (149.00 -- 150.99 GHz), and spw3 (151.00 -- 152.99 GHz), for a total usable continuum bandwidth of 7.5 GHz. Each spectral window contains 128 channels, each 15.625 MHz wide. Higher frequency (Band 7) observations were also approved but, unfortunately, not obtained due to higher priority demand.

The data were calibrated using the ALMA Science Pipeline. To create the Band 4 continuum image, all channels containing line emission were excised out from the calibrated visibilities and the remaining channels were then averaged together, using the CASA task \texttt{split}. The channels containing line emission were determined from sharp peaks in a plot of channel vs. amplitude. To prevent the diffuse clouds in NGC 2264 from obscuring the emission from KH 15D, coverage from the shortest baselines ($<$40 k$\lambda$) was excluded from the data set during this step. The 40 k$\lambda$ value was chosen to minimize the rms noise in the continuum image, providing the best balance between reducing the flux contribution from the extended cloud emission while not excising so much data that the noise increased due to missing visibilities. Note that a 40 k$\lambda$ baseline corresponds to a cutoff of spatial scales greater than 5.2$\arcsec$ ($\sim$ 4000 AU), and thus we do not anticipate any loss of sensitivity to extended KH 15D disk emission. This technique reduced the rms noise of the image from 220 $\mu$Jy to 46 $\mu$Jy. Deconvolution was carried out using the CLEAN algorithm and applying natural weighting. The size of the synthesized beam is $1.4\arcsec \times 1.8\arcsec$.

In addition to the continuum image, channel maps were created for each of the two spectral features present in the visibilities (details on these spectral lines will be discussed in Section \ref{section:alma_spectral_lines}). The continuum emission was subtracted using the CASA task \texttt{uvcontsub}, and again the CLEAN algorithm (natural weighting) was used to deconvolve the channel maps from the dirty beam. The velocity resolution in both ALMA channel maps is coarse (68 km/s in spectral window 0, 62 km/s in spectral window 2) since the original science goal was to attempt to detect the source in the Band 4 continuum image.

\section{Results}

\subsection{Optical Photometry and Stellar Characteristics} \label{section:stellar_characteristics}
\begin{figure}[h!]
\centering
\epsscale{1.0}
\plotone{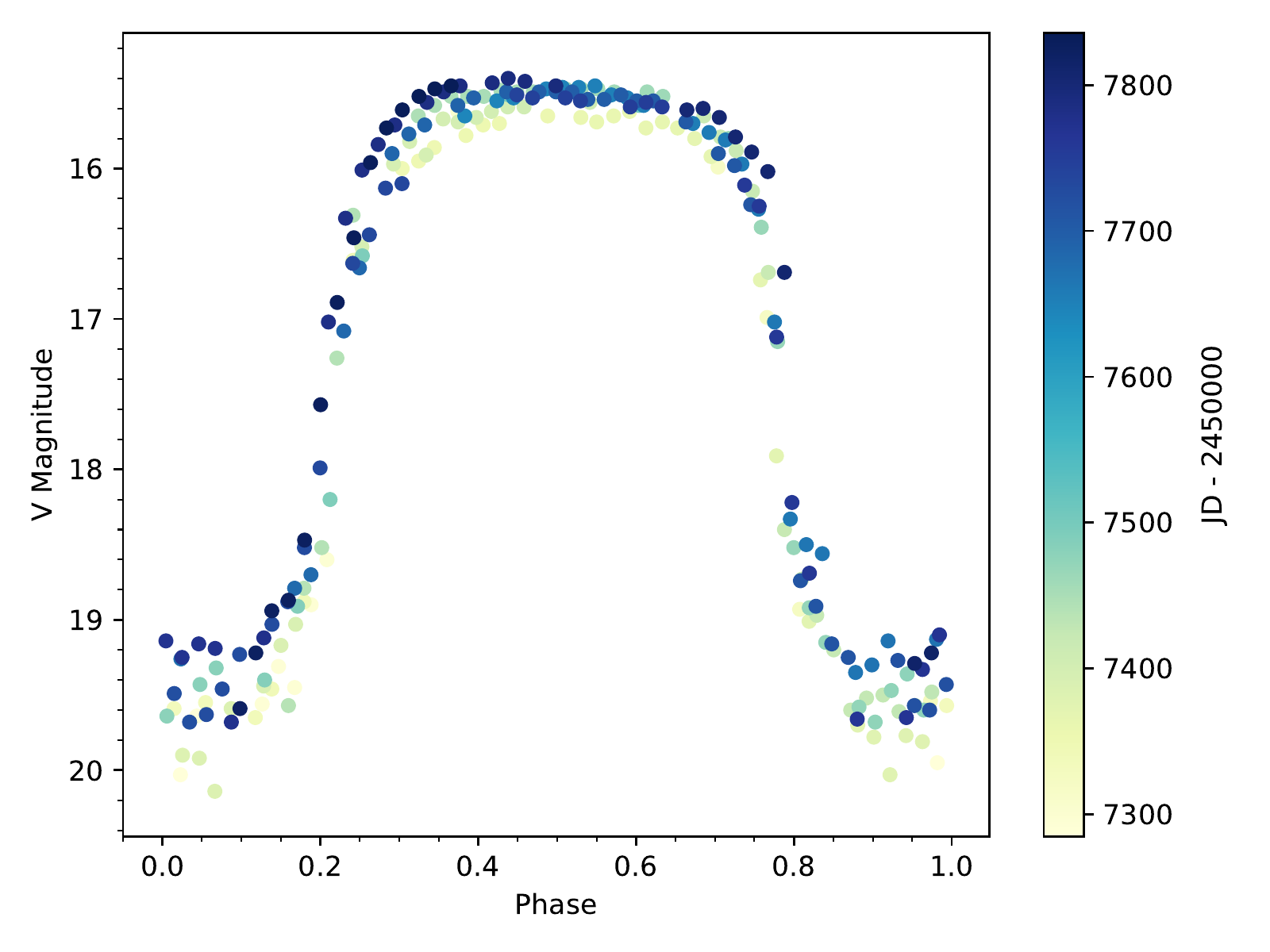}
\caption{Phase-folded V band photometry of KH 15D from the last two observing seasons. While the time spent near maximum brightness continues to increase, the brightness at maximum light has leveled off. We have used the brightness measurements in the 2016/17 observing season (JD $>$ 2457600)  between phases 0.35--0.65 to determine the uneclipsed brightness of Star B, as given in Table 3. As noted in the text, brightness variations of 0.1 mag or so can be expected at any phase due to star spots.
\label{figure:phasefold_V}}
\end{figure}

\begin{figure}[h!]
\centering
\epsscale{1.0}
\plotone{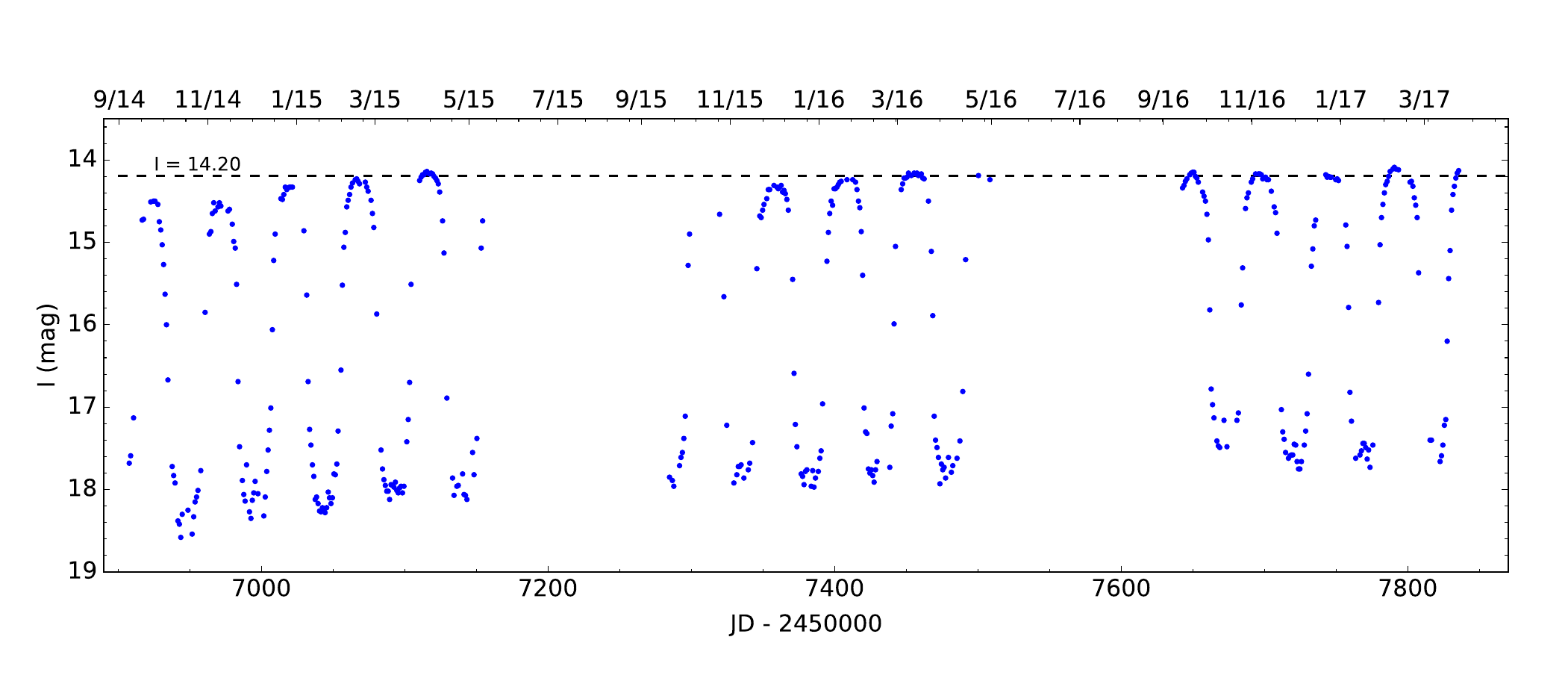}
\caption{Enlargement of the behavior of KH 15D in the I magnitude over the last three seasons. The adopted value of I = 14.20 for the average unocculted brightness of Star B is shown. It seems clear that the star is now fully emerging from behind the occulting ring at each apastron passage. It may also be seen that there was not a perfectly steady rise to this state, indicating that there are either irregularities in the smoothness of the ring edge or that the spottedness of the star affects its brightness substantially, or both.
\label{figure:lightcurve_weirdness}}
\end{figure}

In this new era for KH 15D, represented schematically in panel 4 of Figure \ref{figure:intro_figure}, the entirety of Star B is visible above the disk at each apastron passage, while Star A remains occulted at all phases of the binary orbit. We, therefore, wished to measure for the first time the mean brightness and color of the fully unocculted Star B. To do this we selected the flat portions of the light curve peaks in the 2016/2017 data, corresponding to light curve phases between 0.35 and 0.65 (see Figures \ref{figure:phasefold_V} and \ref{figure:lightcurve_weirdness}). Zero phase corresponds to the mid-point of the eclipse. A similar sample for Star A was generated using data from July 1998 through May 2005, when Star A was the lone visible component, selecting orbital phases between 0.31 and 0.70. The differences in phase between the two samples is due to the slightly different positions of the disk edges in relation to the projected orbit of the stars (leading edge for Star A, trailing edge for Star B). The resulting mean magnitude of each component in VRIJHK and the standard deviation of the mean is given in Table \ref{table:stellar_mags}. These represent our best estimates of the mean unocculted apparent brightness of each component, uncorrected for any scattered light which may be present. The I value for Star B (14.20) is in excellent agreement with the prediction of \citet{WindemuthHerbst2014} of I = 14.19 $\pm$ 0.06 based on the brightness of Star A and the archival photographic data (see below). 

Each star is likely to be variable at about the 0.1 mag level, as they are spotted, rotating T Tauri stars. In fact, the rotation period of Star A, 9.6 days, has been determined from its periodic variations seen during unocculted time intervals \citep{Hamilton2005}. This could account for most or all of the scatter seen in the magnitudes at maximum brightness of each star. We have searched for a rotation period for Star B but the time interval that the star is definitely unocculted is currently still too short to be certain of any result. We do find a hint of a periodicity near 10 days. That would be interesting because, as \citet{Herbst2005} have discussed, there is evidence that Star A has been pseudo-synchronized into its current rotation period through tidal interaction with Star B during periastron passages. If so, one might expect Star B to also be pseudo-synchronized with about the same period. No attempt has been made to identify the most ``pristine" state of the star, i.e. when it is least spotted. We have simply averaged together all of the data during the unocculted phases, forming an ``average" brightness of the photosphere, representing a typical spot coverage, whatever that might be. 

Another factor influencing these results is the possible fine structure, or corrugation, of the disk edge \citep{Herbst2002}. This complicates a bit the task of identifying when, exactly, the star is unobscured. As may be seen from Figure \ref{figure:lightcurve_weirdness}, there has not been a completely steady progression of Star B towards its maximum brightness. In 2014/15 the star appeared to reach full visibility during the last cycle of that season, but in 2015/16 it had apparently become slightly occulted once again during the first cycle and progressively rose in brightness during that observing season, mimicking its 2014/15 behavior. This could be due to the irregular nature of the disk edge (i.e. some hills and valleys) as discussed for Star A by \citet{Herbst2002}. Or it could involve the spottedness of the star, especially if there is pseudo-synchronization into a 5:1 spin orbital resonance, as suggested by \citet{Herbst2005}. Most likely, both effects contribute to the brightness pattern. Resolution of these issues must await the progression of the disk edge to the point where Star B is clearly unocculted for a sufficiently long period of time to study its rotational modulation. 

\begin{figure}[h!]
\centering
\epsscale{0.8}
\plotone{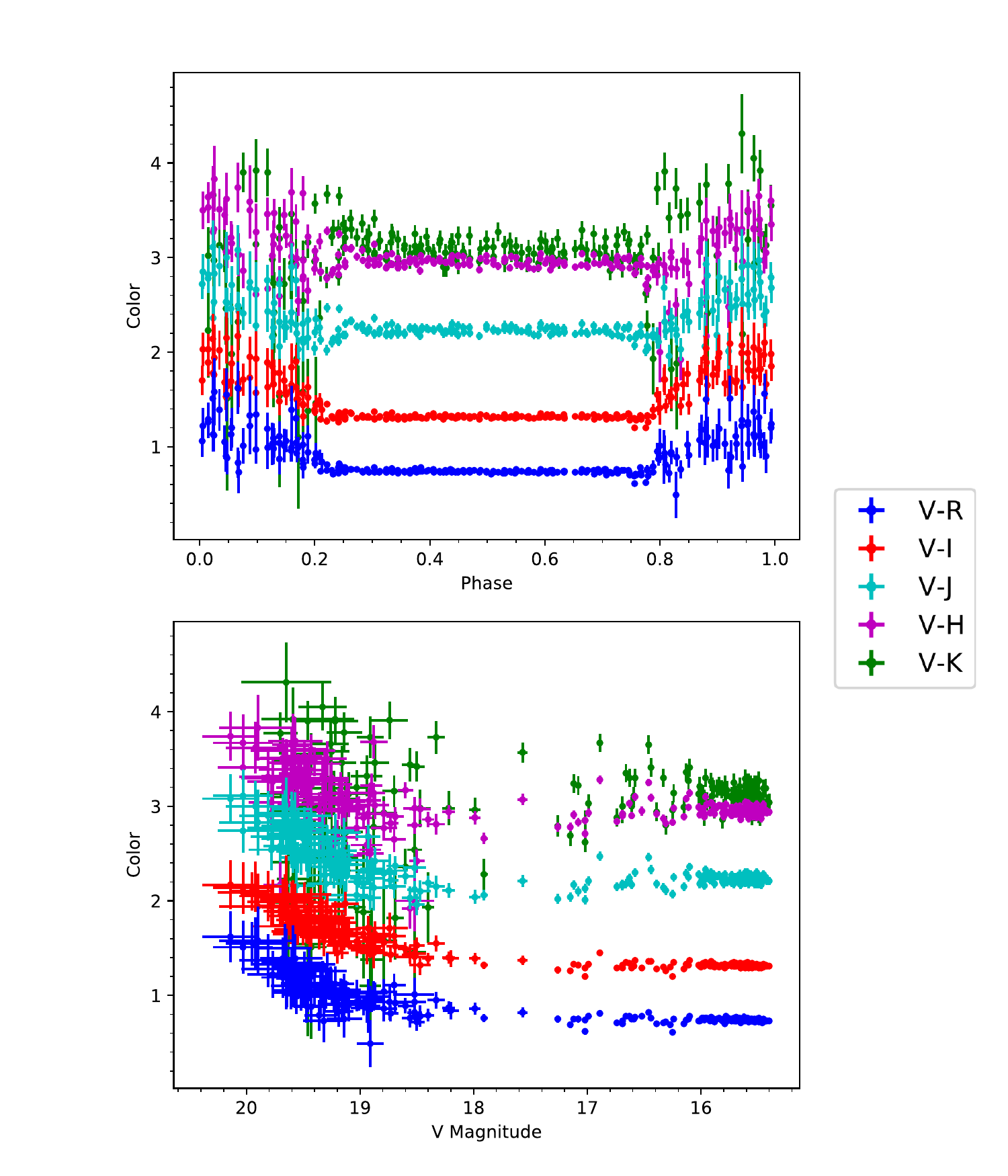}
\caption{Color vs. orbital phase (top) and color vs. V magnitude (bottom) over the last two observing seasons (JD $>$ 2457285). During the bright phase, the system's color is constant within 0.1 mag for V-R and V-I, and within 0.2 mag for V-J and V-H. As the system fades, it first becomes bluer due to more efficient scattering at short wavelengths \citep{Nicole}. When both stars are obscured during the faint phase, the system becomes redder, and there is more scatter due to lower photometric precision. The extreme redness may be due to third light in the system from a brown dwarf or luminous giant planet \citep{WindemuthHerbst2014,Nicole2}.}
\label{figure:color_plots}
\end{figure}

In Figure \ref{figure:color_plots}, we plot the color of the system over the last two observing seasons against both its orbital phase and V magnitude. Near the system's bright phase (V mag $<$ 15.6, phase between 0.35--0.65; see Figure \ref{figure:phasefold_V}) Star B's unobscured photosphere dominates the total magnitude, and the system's brightness and color is independent of phase. The new data are consistent with the overall picture described by \citet{Nicole}, who included longer wavelength (Spitzer) data in their analysis. As is discussed in that paper, the system first gets bluer as it fades (an effect most easily seen here in the V-I and V-J data), due to the wavelength dependence of the scattering phase function. When it is very faint, the system becomes extremely red, as is clearly seen in Figure \ref{figure:color_plots}. This indicates the likely presence of a third, low temperature, light component in addition to stars A and B. \citet{WindemuthHerbst2014} and \citet{Nicole2} argue that this could be light from a brown dwarf or luminous giant planet component.

The mean unocculted I magnitudes for Star A (14.49) and Star B (14.20) are plotted on Figure \ref{figure:intro_figure} and may be compared with expectation based on the photographic photometry record that extends back to the time when both stars were visible together \citep{JohnsonWinn2004,Johnson2005,Maffei2005}. Note that there may be a scattered light component to each of the stars' individual magnitudes, which could lead to overestimates of their luminosities. Until a more comprehensive scattering model is devised for the system it is hard to know how to handle that potential contribution. Based on the individual stellar brightnesses determined here, ignoring any scattered light component, we would predict the brightness of the system to be I = 13.58 at times when both stars were unobscured. That value is shown in Figure \ref{figure:intro_figure}. The last time both components were visible simultaneously and (photographic) photometry was obtained was during the period 1951--1991. Using the available archival photographic plate data from \citet{JohnsonWinn2004}, \citet{Johnson2005}, and \citet{Maffei2005}, \citet{Winn2006} determined the mean brightness of the combined starlight at this time to be I = $13.57\pm0.03$ mag, in excellent agreement with our result. There are severe difficulties in transforming from the photographic magnitude system to the CCD system in use today and the authors of the archival work have noted that one might anticipate an uncertainty in the zero point magnitude of $\sim 0.15$ mag \citep{JohnsonWinn2004}. Hence, the near perfect agreement between the summed CCD magnitudes and the historical photographic photometry may be somewhat fortuitous. However, it does indicate that the transformation to Cousins-I was successful and provides further support for our overall interpretation of the light curve of KH 15D. 

To derive physical parameters for the stellar components we need to adopt a distance and reddening for KH 15D. Table \ref{table:cluster_facts} provides a summary of all the published distance measurements to NGC 2264, the cluster of which KH 15D is a member, that we could find. In most of these studies, the distance was determined by spectrophotometric fitting of main sequence stars; exceptions are mentioned in the table notes. Based on this set of studies, we adopt a distance of 800 $\pm$ 20 pc (the unweighted mean, omitting the discrepant \citet{Dzib2014} value). The mean foreground reddening to the cluster is only E(B-V) = 0.075 $\pm$ 0.003, as recently derived by \citet{Turner2012}. There is relatively little differential reddening within the cluster and a comparison of the uncorrected colors and spectral types for Stars A and B suggests that, like its cluster neighbors, KH 15D experiences relatively little foreground extinction. We have adopted a reddening value of E(B-V) = 0.075 and a standard extinction law from \citet{Cardelli1989}, which has a ratio of total to selective extinction of R $=3.1$. The resulting extinction corrections and derived absolute VRIJHK magnitudes of both stars out of eclipse are presented in Table \ref{table:stellar_mags}. The intrinsic V-I color was then used to estimate the effective temperature and bolometric correction, BC$_\mathrm{V}$, of each star following Table 6 of \citet{PM13}. Stellar luminosities were calculated from the resulting bolometric magnitudes, and radii were determined from the effective temperatures and luminosities. The full set of derived properties for each star is given in Table \ref{table:derived_stellar_props}. The errors listed in Table \ref{table:derived_stellar_props} result from the 1$\sigma$ error on V-I, which corresponds to a spread in spectral type of approximately $\pm$1 subclass. Masses and ages are, of course, model dependent as discussed in what follows.

\subsection{Radio Continuum Results and a Disk Mass Upper Limit}
The main goal of the radio observations was to detect the inferred KH 15D accretion disk directly through its expected thermal continuum emission from millimeter-sized grains present in the outer disk. At the outset we knew this would be difficult because of the distance to the system and the confusing region, full of diffuse clouds, in which it is embedded. Scaling from disk surveys of nearby star forming regions, we predicted a less than 50-50 chance that accretion disk emission would be detected for this WTTS. However, because of the importance of the object, discussed above, we felt it was worth the attempt.

\subsubsection{SMA Continuum Results}
The first submillimeter attempt to target KH 15D was the 2010 Oct 13 SMA filler track. This yielded a tenatative detection of $5.1\pm1.9$ mJy, by using the miriad task \texttt{uvfit} to fit a point source model to the visibilities. Imaging revealed the peak in flux density to be centered at the location of KH 15D. The subsequent observation with the SMA on 2011 Oct 25, with higher sensitivity, gave a similar low significance result using a point source model fit to the visibilities, but did not show a clear peak associated with KH 15D in the image. Taken together, these uncertain results spurred submission of a proposal to ALMA for much more sensitive continuum observation aimed at confirmation.

\subsubsection{ALMA Non-Detection}
ALMA observations in Band 4 (1959--2188 $\mu$m; 137--153 GHz) did not reveal a source at the location of KH 15D. We may use the non-detection to set an upper limit on the flux from the disk, and a constraint on its mass. The final cleaned ALMA Band 4 image has an rms noise of $\sigma=$ 46 $\mu$Jy, with no significant emission at the coordinates of KH 15D. We can therefore place a 3$\sigma$ upper limit on the 2 mm flux density of the disk at 0.14 mJy. Note that we do not expect the disk to be spatially resolved, as the angular resolution of the data (1.5$\arcsec$) corresponds to a linear scale of $\sim$1200 AU at the distance of KH 15D. A non-detection is a little surprising given the 4$\sigma$ detection with SMA, at lower sensitivity. Possible explanations for this apparent contradiction will be examined in Section \ref{section:compare_radio_cont}. 

If we assume that the emission is from an isothermal and optically thin portion of the disk, we can estimate an upper limit for the total mass of the disk using the following relation \citep{AndrewsWilliams,Hildebrand83}:
\begin{equation} 
M_d = \frac{d^2 F_\nu}{\kappa_\nu B_\nu(T_c)}
\end{equation}
where d is the distance to the source, $F_\nu$ is the flux density at the observed frequency $\nu$, $\kappa_\nu$ is the opacity, and $B_\nu(T_c)$ is the Planck function at a characteristic temperature $T_c$. The opacity was estimated to be $\kappa_\nu \approx$ 3 (1 mm$/\lambda)$ cm$^2$ g$^{-1}$. This approximation was derived by Beckwith \& Sargent (1991) and assumes grains are between 0.5 and 20 $\mu$m in size. A characteristic temperature of 25 K \citep[the approximate mean temperature in a survey of disks in Taurus-Auriga, conducted by][]{AndrewsWilliams} was chosen, resulting in an upper mass limit of 0.0016 $M_\odot$, corresponding to 1.7 Jupiter masses. This mass estimate is sensitive to assumptions about the opacity and characteristic temperature. Further, we can only detect particles smaller in size than the wavelength of observation. Note, however, that this is still fairly large for a disk mass; the upper limit is larger than 55\% of the known disk masses in the \citet{AndrewsWilliams} survey. 

\subsection{Radio Line Emission: a Molecular Outflow and Background Cloud Emission}

While the radio observations were designed to search for thermal emission from the accretion disk, which was not detected, they also permitted a search for molecular line emission, which {\it was} detected in both data sets. The SMA data revealed a CO J=3--2 line source that is spatially coincident with the filamentary jet of shocked gas seen in the near-IR and very likely associated with the KH 15D system, if not the binary itself. The ALMA data revealed two spectral features that probably arise within the diffuse cloud emission in NGC 2264, but appear unrelated to KH 15D.

\subsubsection{SMA Line Emission}
\begin{figure}[h!]
\centering
\epsscale{1.0}
\plotone{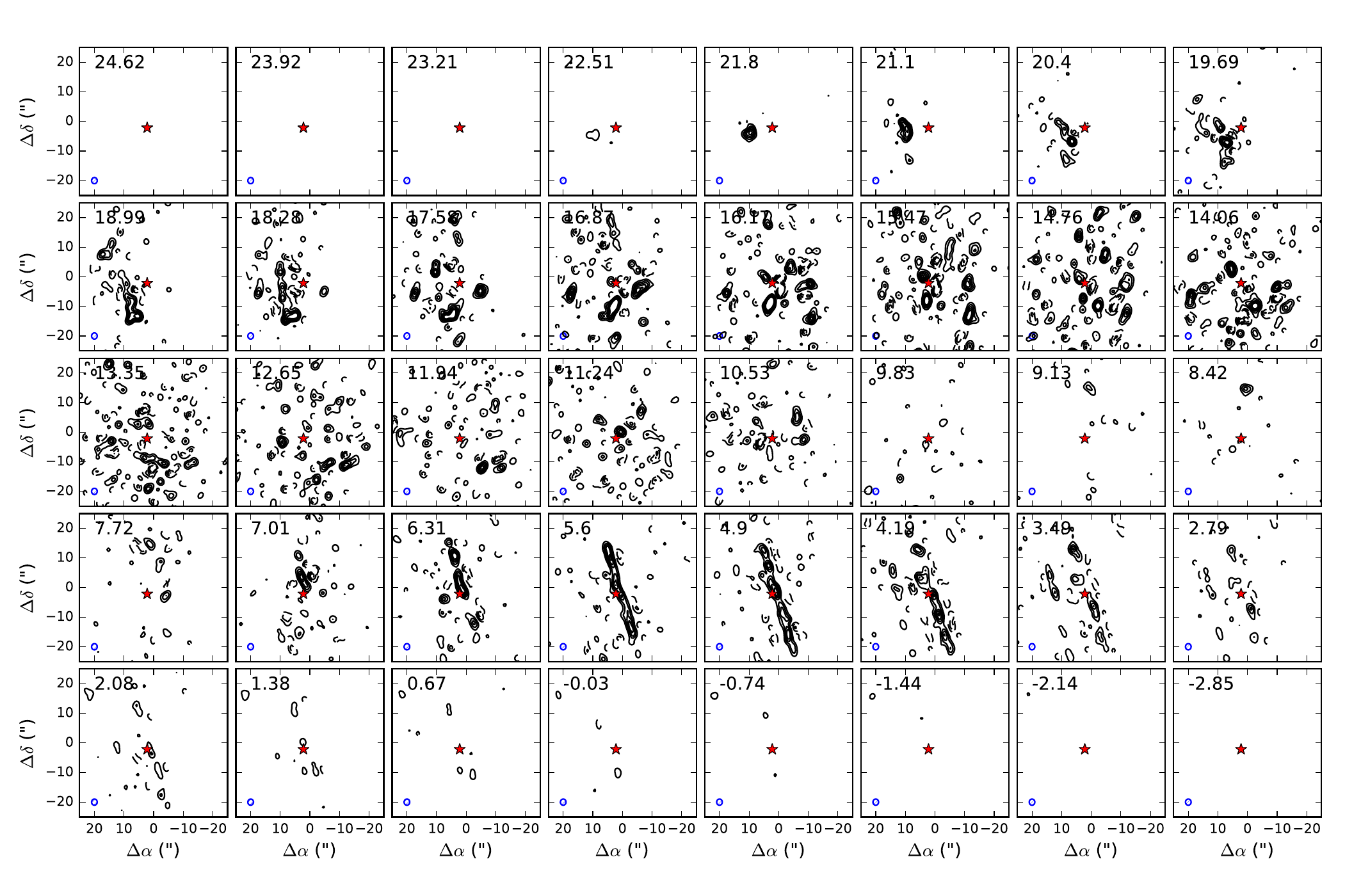}
\caption{Channel maps showing the CO J=3--2 emission. For each channel, the central LSR velocity (km/s) is given in the top left, the synthesized beam is indicated in blue in the bottom left, and the red star represents the position of KH 15D. Contours are [-10, -8, -6, -4, -2, 2, 4, 6, 8, 10] x 0.21 Jy/beam. The emission is dominated by the ambient cloud between 10--21 km/s. Evidence of an extended molecular outflow is found around the position and systemic velocity (3.4 km/s) of KH 15D.
\label{figure:full_channel_map}}
\end{figure}

The SMA velocity channel maps of the CO J=3--2 (Figure \ref{figure:full_channel_map}) reveal a detailed image of the ambient cloud emission throughout the region. The clouds emit across kinematic local standard of rest (LSR) velocities of 10--21 km/s. Additionally, there is a linear structure present in the blue-shifted half of the CO line, at 3--7 km/s. These maps suggest a spatial and kinematic association between the structure and KH 15D. They align perfectly with the filamentary jet of shocked H$_2$ emission discovered by \citet{Tokunaga2004} and span a velocity range close to the binary's systemic velocity (3.4 $\pm$ 1.3 km/s) inferred from the radial velocity measurements of Star A \citep{Johnson2004} and a model of the system by \citet{Winn2006}. Thus we believe that this structure is associated with the KH 15D system, and that it could be a detection of the binary outflow first proposed by \citet{Hamilton2003} and discussed in more detail by \citet{Mundt2010}.

\begin{figure}[h!]
\centering
\epsscale{1.0}
\plotone{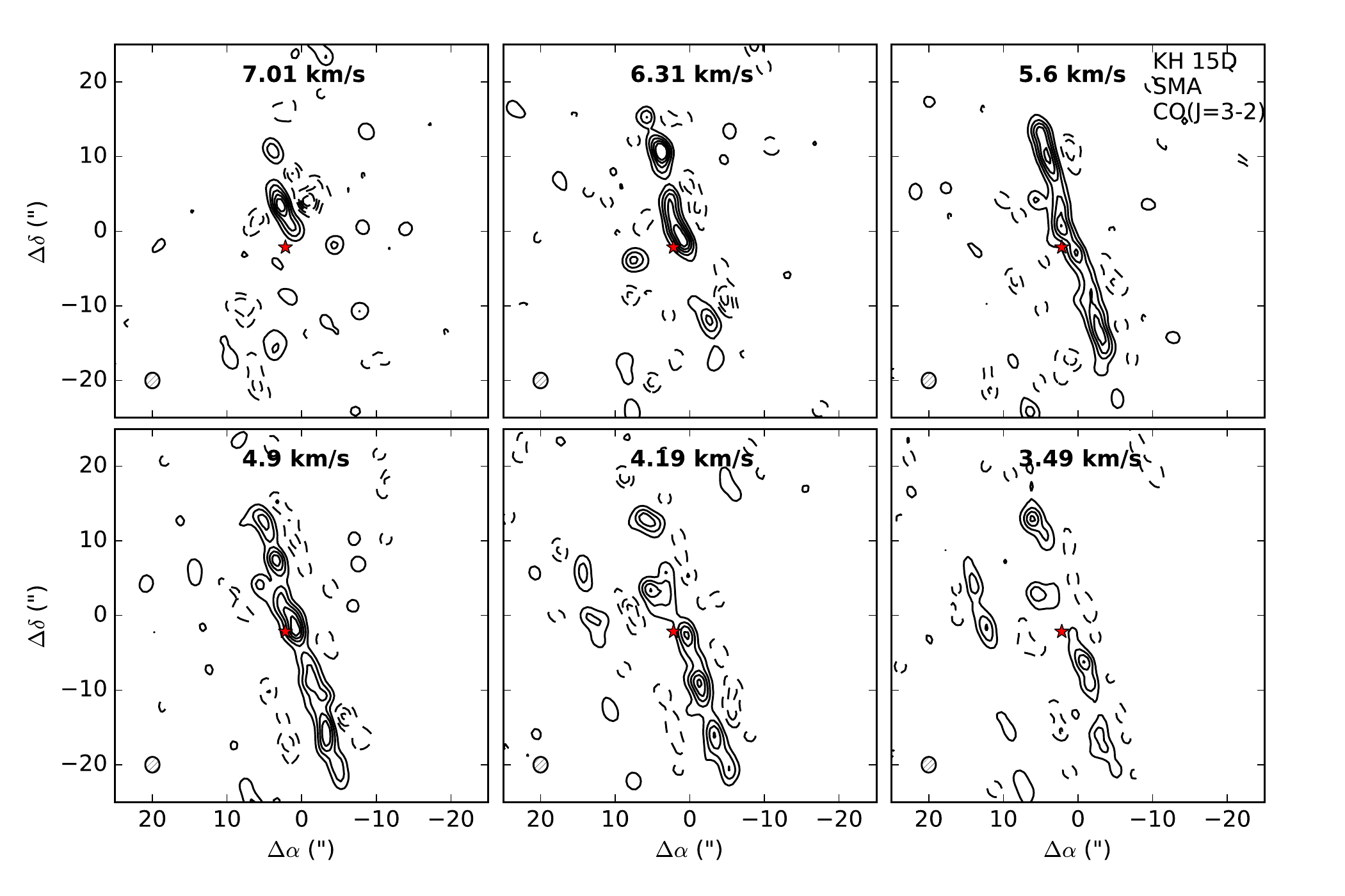}
\caption{Six channels (0.7 km/s in width) showing the extent of the CO J=3--2 emission associated with the jet. Contours are [-10,-8,-6,-4,-2,2,4,6,8,10] x 0.21 Jy/beam. The northern half is slightly redshifted while the southern half is slightly blueshifted. The low velocities imply that the inclination is small, and that the jet is oriented nearly perpendicular to the line of sight. Note that the origin of the SMA image (6:41:10.18 +09:28:35.50) is slightly offset from the known optical/IR position of KH 15D, taken to be the mean of its optical position as reported in SIMBAD (6:41:10.31, +9:28:33.2, J2000) and its Spitzer infrared position (6:41:10.34, +9;28:33.5, J2000). It is source ID 462852 in the CSI 2264 program, also known as Mon 1370 \citep{Cody,Nicole}. 
\label{figure:COmap}}
\end{figure}

A close-up of the channels containing emission from the molecular outflow is shown in Figure \ref{figure:COmap}. The northern half is slightly redshifted compared to the southern half. Together with the low radial velocities of the CO emission \citep[jet speeds are typically 100--200 km/s;][]{Mundt1983}, this implies that the jet is oriented nearly perpendicular to the line of sight, with the southern portion inclined slightly towards the direction of the observer. This reinforces the jet inclination angle of 84 degrees proposed by \citet{Hamilton2003}.

In Figure \ref{figure:jet_overlap} we overlap the image of the H$_2$ emission from \citet{Tokunaga2004} over the CO emission integrated across the channels exhibiting jet-like structure. The images are aligned using an average of the optical (6:41:10.31 +9:28:33.2; SIMBAD) and infrared (6:41:10.34 +9:28:33.5; CSI NGC2264) positions of KH 15D \citep{Cody}. The shape of the emission aligns nearly perfectly along the axis of the outflow. Even the faint nebulosity visible in the southern border of the H$_2$ image, speculated by \citet{Tokunaga2004} to be a re-emergence of the hidden counterjet, aligns with a knot of CO emission. The similarities in the shape and knots of emission between the two linear structures makes the case that they are associated with one another and with KH 15D. There is, nonetheless, an intriguing small, but real, offset ($\sim 1 \arcsec$ or 800 AU) of the stellar image from the central line of the molecular emission, which we discuss further in Section \ref{section:CO_outflow}.  

\subsubsection{ALMA Line Emission} \label{section:alma_spectral_lines}
Two spectral features were detected in the calibrated ALMA visibilities, however neither is localized to the position of KH 15D. The line emission is diffuse and most likely coming from galactic molecular clouds associated with or along the line of sight to NGC 2264. In an attempt to identify these spectral features, we submitted a Splatalogue\footnote{http://www.cv.nrao.edu/php/splat/} search querying all spectral lines with frequencies within the range of the channels displaying line emission. We then restricted our query to molecules that have been previously detected in the interstellar medium or in circumstellar shells using the Molecules in Space database hosted by Universit{\"a}t zu K{\"o}ln\footnote{http://www.astro.uni-koeln.de/cdms/molecules}. The data were then shifted to the rest frequency of each of these candidate spectral features using the CASA task \texttt{cvel}, and a Gaussian was fit to the shifted profile in order to find the central velocity. This velocity was compared to the velocity of the cloud emission in the SMA channel maps (10--21 km/s). The best fit for the spectral feature present in spectral window 0 was SO ($\nu_0$ = 138178.6 MHz). The feature present in spectral window 2 is less certain due to the coarse velocity resolution. This analysis points to either SO$_2$ ($\nu_0$ = 150486.92 MHz), CH$_3$OH ($\nu_0$ = 150497.826), or H$_2$CO ($\nu_0$ = 150498.334 MHz’) as the likely source of the emission. However, the line profile appears to be double peaked and none of these candidate spectral features have any known splitting. Thus we cannot say with certainty the source of the molecular cloud emission. 

\section{Discussion}

\subsection{Stellar Characteristics} \label{section:stellar_characteristics_discussion}

The rise of Star B sets the stage for an improvement to the system model \citep{Winn2006} that should yield highly accurate stellar characteristics for T Tauri stars. There is only one other known low-mass, PMS binary in NGC 2264 with determined stellar parameters (CoRoT 223992193), which also shows evidence of an accretion disk \citep{Gillen2014}. Thus the KH 15D system provides a valuable set of data points for improving pre-main sequence stellar models. As a prelude to the upcoming KH 15D model, we explore what the photometry alone predicts about the stellar components. The full model requires radial velocity information on Star B that has not yet been obtained, as well as a prescription for the scattered light component. Here we see what constraints the data in hand place on the stars independently of those considerations. 

An output of the full model to come will be the stellar radii, which are highly constrained by the shape of the eclipse light curve. In combination with the stellar luminosities, these will yield effective temperatures of high precision. With the data in hand, we reverse this procedure and use derived effective temperatures to predict the stellar radii. For main sequence stars, effective temperatures are usually estimated from their spectral type and/or color. Stars A and B have been reported to be K7 and K1 T Tauri stars, respectively, by \citet{Hamilton2001} and \citet{Capelo2012}. However, these determinations are based on different spectral regions (optical and near-IR, respectively) and correlations between spectral type and effective temperatures for pre-main sequence stars are notoriously hard to quantify, given the complex, magnetically-active photospheres of young stars \citep{CohenKuhi1979, Hillenbrand1997}. Some authors have attempted to adapt the spectral type vs. effective temperature relationship of main sequence stars for young stars. For example, \citet{PM13} proposed effective temperatures for 5-30 Myr stars of 4920 K and 3970 K for K1 and K7 stars, respectively. Other authors, such as \citet{CohenKuhi1979} and  \cite{Hillenbrand1997}, simply adopt dwarf scales and apply larger uncertainties. Given the uncertainties inherent in these methods, as well as the lower precision data based on spectral types, we prefer to use the intrinsic color as our primary guide to effective temperatures. This choice is reflected in Table \ref{table:derived_stellar_props}. 

\begin{figure}[h!]
\centering
\epsscale{1.0}
\plotone{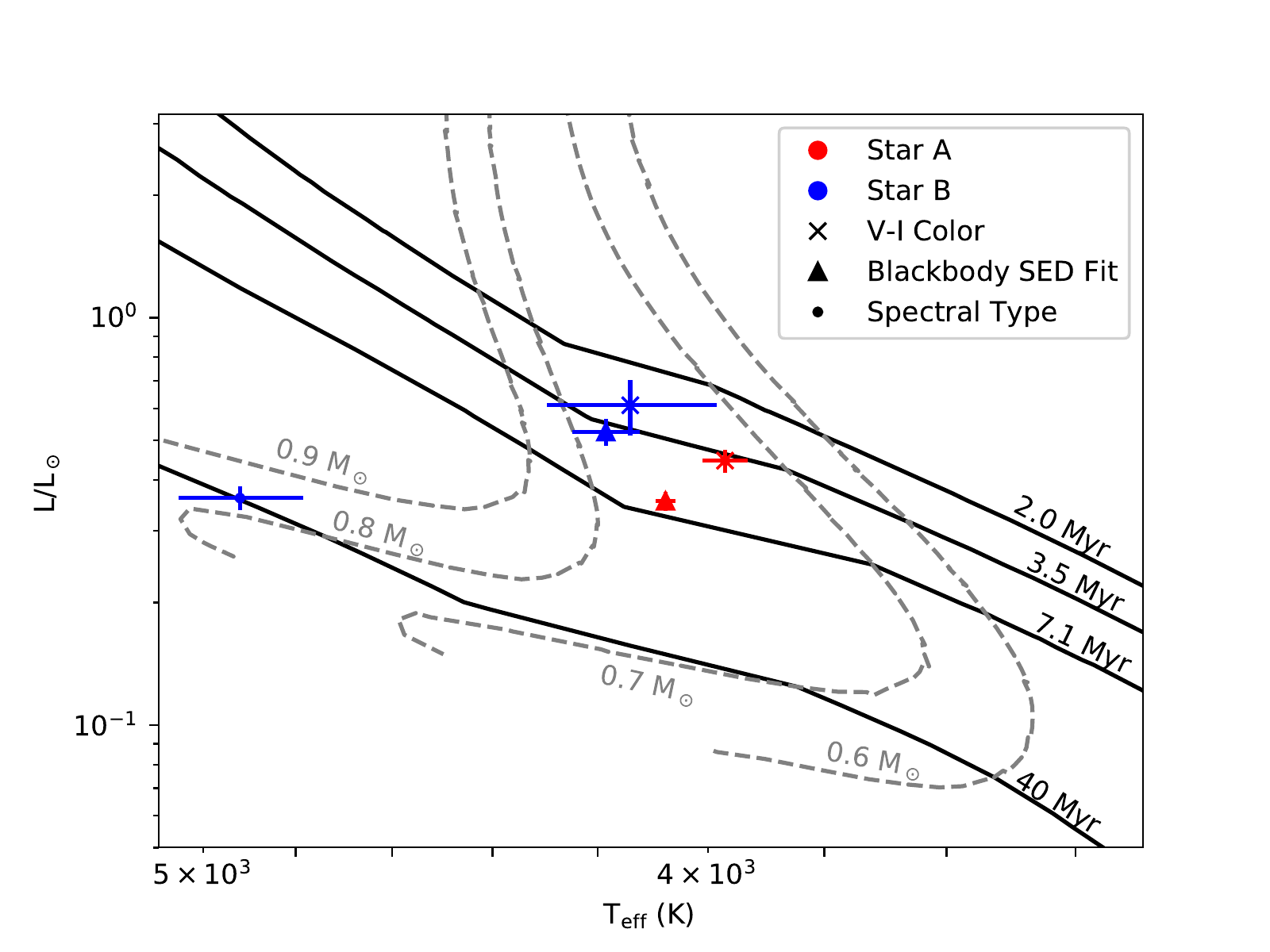} 
\caption{The resulting locations of Star A and Star B on an H-R diagram when using three different methods of estimating effective temperatures for PMS stars. Stellar isochrones are from \citet{Marigo2017}, accessed via \texttt{http://stev.oapd.inaf.it/cmd}. Spectral type estimates of the effective temperature are those discussed in Section \ref{section:stellar_characteristics_discussion} from \citet{PM13} (errors taken to be $\pm$ 1 spectral subclass). Note that for Star A, this estimate overlaps with that of the V-I color method, since its color is consistent with its determined spectral type. Clearly the spectral type is not a reliable method for fitting effective temperature in this circumstance, since a common age cannot be matched to both stars in the system. The blackbody SED fit method is also slightly problematic, since it produces model-dependent ages of the stars that differ by $\sim$3 Myr, and can only be fit by a single age (5.0 Myr) at the 3$\sigma$ level. The 3.5 Myr isochrone fits both data points derived using the V-I color of the stars within the 1$\sigma$ level.
\label{figure:isochrones}}
\end{figure}

The stellar radii presented in Table 5 (1.41 $\pm$ 0.05 R$_\odot$ for Star A, 1.52 $\pm$ 0.16 R$_\odot$ for Star B) suggest that the two stars are about the same size. This appears sensible; the stars are presumably the same age, and the relatively small difference in brightness and color between the components, along with the \citet{Winn2006} model result, points to roughly equal masses (with Star B being slightly brighter, bluer, and therefore more massive). Using the \citet{PM13} calibration, the V-I colors imply spectral types of K7 and K5, respectively, for Stars A and B. This result is consistent with the spectral type determination of K7 by \citet{Hamilton2001} based on optical spectra. It is marginally inconsistent with the spectral type of K1 determined by \citet{Capelo2012} based on near-IR spectra. This may simply reflect the difficulties of assigning consistent spectral types to T Tauri stars. If the \citet{PM13} effective temperature and bolometric correction corresponding to spectral type K1 are used in place of those determined from the mean V-I color of Star B, the models give ages that differ by 36 Myr for the two stars in the system (using PMS evolutionary tracks from \citet{Marigo2017}; see Figure \ref{figure:isochrones}). This reinforces our decision to use the stars' colors instead of their spectral types for the effective temperature estimates. It would be of interest to obtain additional optical spectra of Star B while it, alone, dominates the spectrum of KH 15D, for the purpose of obtaining a better spectral classification. 

Alternatively, the effective temperatures can be estimated by fitting a blackbody curve to the full VRIJHK spectral energy distribution of each star. As in the V-I estimate, we adopted the mean reddening to NGC 2264, E(B-V)=0.075, from \citet{Turner2012} and the extinction law from \citet{Cardelli1989}, yielding A$_\mathrm{V}=0.23$. For each fit, the stellar angular diameter was allowed to vary as a function of the effective temperature and measured stellar magnitude, using the bolometric correction calibration from \citet{Hillenbrand1997}. This results in T$_{\mathrm{eff}}=4076\pm18$ K, R $=1.19\pm0.03$ R$_\odot$ for Star A and T$_{\mathrm{eff}}=4185\pm63$ K, R $=1.38\pm0.07$ R$_\odot$ for Star B.  The errors on the temperatures are representative of the errors on the fit, and do not encompass the much larger systematic errors such as the assumption that each star is a perfect blackbody. The V-I color approach was favored because of the smaller systematic errors, which makes it easier to properly quantify our (large) uncertainties. Additionally, the temperatures derived from the color provide a nicer fit to the \citet{Marigo2017} isochrones (see Figure \ref{figure:isochrones}). As noted previously, future dynamical models of the KH 15D system will derive the radii of the components to remarkable accuracy for PMS stars  (within 10\%). This presents the opportunity to work backwards to derive the effective temperatures, creating valuable data points for better characterizing pre-main sequence stars in the future. 

The effective temperatures of the binary components, derived by each of the three approaches described above, are plotted against their corresponding luminosities on an HR diagram in Figure \ref{figure:isochrones}. Also plotted are isochrones from \citet{Marigo2017}, which build off of the Padova-Trieste Stellar Evolution Code (PARSEC) previously edited by \citet{BressanIsochrones}. A metallicity of $Z=0.0152$, $Y=0.2485+1.78Z$ (the solar values adopted in \citet{Marigo2017}) is used. The (unweighted) mean metallicity of NGC 2264 is $[\mathrm{Fe}/\mathrm{H}]=-0.08$ \citep{Paunzen2010} and so we find it reasonable to adopt a solar metallicity. The age of the system is likely between 2.5--5.0 Myr, with 3.5 Myr providing the best simultaneous fit to both stars. We note that this representative age range does not reflect any systematic effects that may be present due to inaccuracies of the models or transformations from measured to physical variables. The \citet{Marigo2017} stellar models imply masses of 0.71--0.72 M$_\odot$ for Star A and 0.70--0.83 M$_\odot$ for Star B (Table \ref{table:derived_stellar_props}). As described in more detail in the caption of Figure \ref{figure:isochrones}, the spectral type and blackbody SED fit methods do not produce stellar parameters that can be fit by a common age above the 3$\sigma$ level.

\begin{figure}[h!]
\centering
\epsscale{1.0}
\plotone{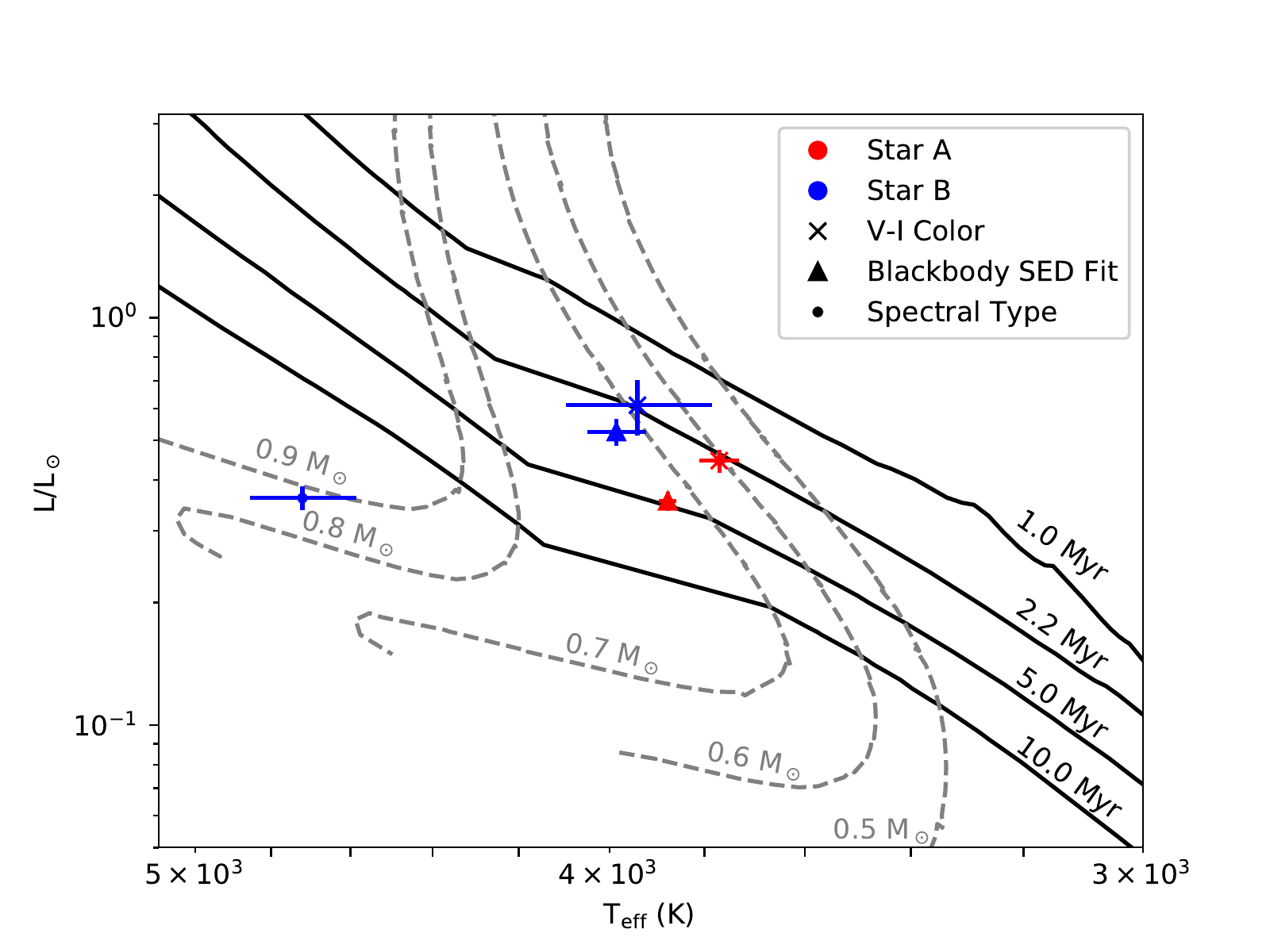}
\caption{Same as Figure \ref{figure:isochrones}, but the effective temperatures of Stars A and B have been shifted by 200 K towards cooler effective temperatures, as discussed in Section \ref{section:stellar_characteristics_discussion}. Again, the blackbody SED fit and spectral type methods do not produce stellar parameters that can be fit to a single age above the 3$\sigma$ level. However the temperature shift does result in a better fit for the age of the system ($\sim$2.2 Myr) using the V-I color method. This method yields mass estimates of 0.60$^{+0.05}_{-0.03}$ M$_\odot$ for Star A and 0.68$^{+0.05}_{-0.13}$ M$_\odot$ for Star B.
\label{figure:isochrones_shifted}}
\end{figure}

Effective temperature scales are model-dependent, and have been shown to overestimate temperatures by about 200 K for pre-main sequence stars with masses less than 1 M$_\odot$ \citep{David2016,Kraus2015}. A temperature suppression of this magnitude can be explained by the exclusion of starspots and magnetic fields from the stellar models \citep{Feiden2016,Somers2015}. To address this concern, we shift the measured KH 15D stellar effective temperatures by 200 K towards cooler temperatures, and again plot our results alongside the \citet{Marigo2017} isochrones in Figure \ref{figure:isochrones_shifted}. The shifted plot suggests a slightly younger age of 1.7--4.0 Myr for the system, as well as smaller masses (0.57--0.65 M$_\odot$ for Star A and 0.55--0.73 M$_\odot$ for Star B). The discrepancy between the shifted and non-shifted data demonstrates the effect that systematic uncertainties in PMS star effective temperature scales have on our ability to accurately estimate physical parameters. The shifted values appear slightly more sensible in the context of the PARSEC models than their un-shifted counterparts. Nonetheless, we proceeded in our analysis with the original measurements.
 
Our values of radius and luminosity for Star A are entirely consistent with those determined by \citet{Hamilton2001} and adopted in the \citet{Winn2006} model. The Star B parameters also agree with those derived by \citet{Winn2006}, although we note that the error bars in both studies are substantial. The effective temperature measurement continues to be of interest, since it is the most difficult parameter to determine just from photometric and spectroscopic data. The M$_\mathrm{B}$/M$_\mathrm{A}$ ratio predicted by the Winn et al. model (M$_\mathrm{B}$/M$_\mathrm{A}$ = 1.2 $\pm$ 0.1) is slightly larger than the mass ratio we determined here using the \citet{PM13} effective temperature scale, but the discrepancy is only at the 2$\sigma$ level. Use of the shifted temperature scale reduces the minor discrepancy to within 1$\sigma$. We expect the updated dynamical model to strongly constrain the radii and, thereby, the effective temperatures of these stars. With that anticipated precision, Stars A and B of the KH 15D system should be extremely useful for testing models of pre-main sequence evolution.

\subsection{The ALMA Non-detection} \label{section:compare_radio_cont}

It was disappointing and slightly surprising not to detect thermal emission from the KH 15D accretion disk with ALMA, since the SMA data suggested a flux density which would be easily seen. We are not sure how to reconcile these facts but note the following: 1) The SMA detection did not reach the usual 5$\sigma$ level that traditionally signifies a solid detection. 2) The frequencies of the SMA and ALMA observations were different, but it would require a physically unrealistic spectral index (around $\alpha =$ 4.4) to account for the ALMA non-detection on this basis. 3) Perhaps the SMA detection was of non-thermal emission associated with the source. Some WTTS are non-thermal radio emitters; synchrotron emission from events in their magnetospheres can cause detectable radio outbursts \citep{Bieging}. The stars in KH 15D probably pass close enough to one another at perihelion that their magnetospheres could interact, resulting in ``pulsed" eruptions as are seen in DQ Tau \citep{Tofflemire}. To check on this possibility, we looked at the orbital phase distribution of the radio measurements. Unfortunately for this idea, it turns out that both the ALMA and 2011 SMA data were taken during an optically bright phase of the system near aphelion, whereas the 2010 SMA data was at an intermediate brightness phase. In short, we have no explanation for the apparent detection of KH 15D with SMA and clear lack of detection with ALMA. Obviously it would be of interest to repeat the SMA measurements at some point in the future, with the benefit that it may soon be possible to detect changes in the filamentary CO outflow.

\subsection{The CO Outflow} \label{section:CO_outflow}

\begin{figure}[h!]
\centering
\epsscale{1.0}
\plotone{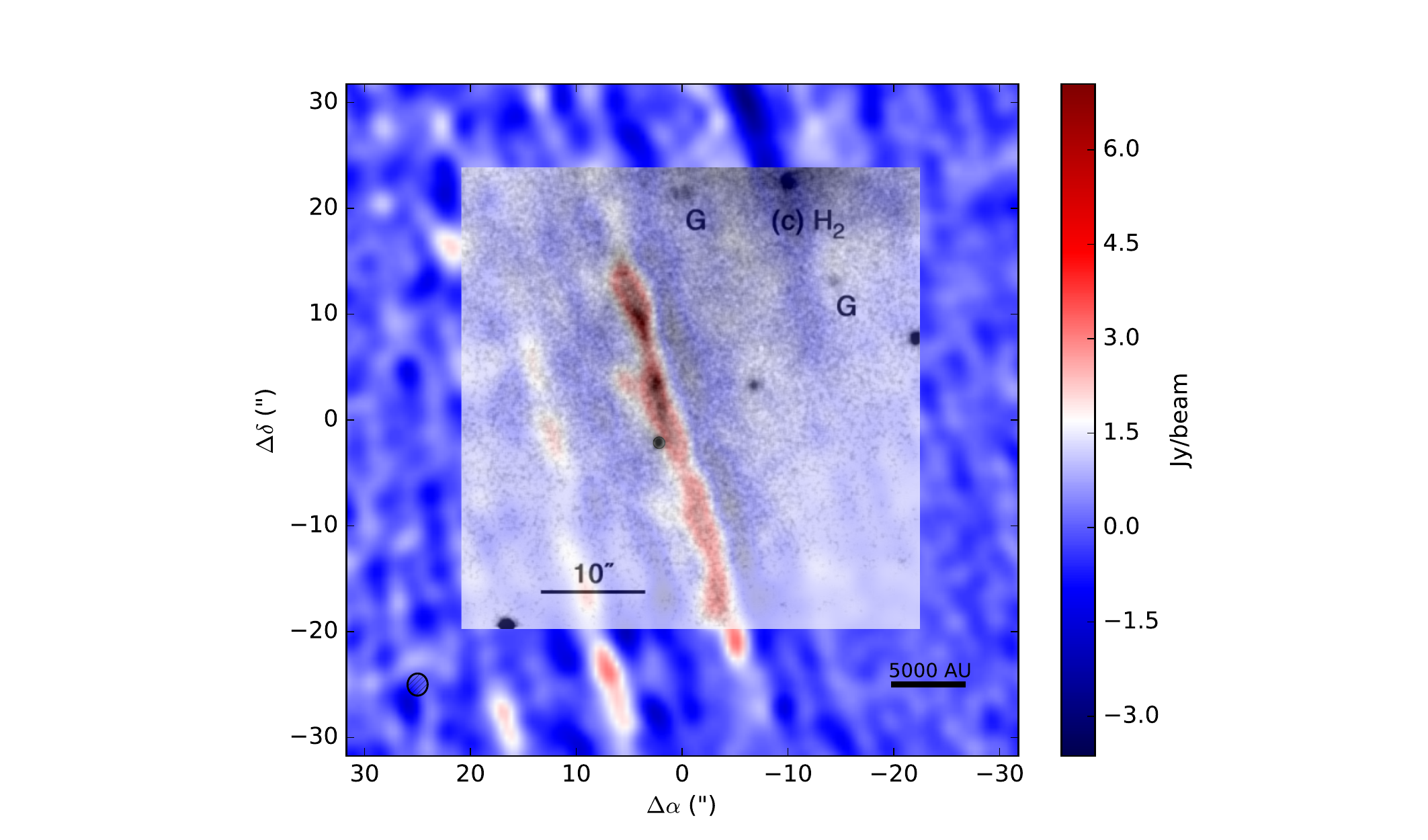}
\caption{The H$_2$ narrow-band image of KH 15D from \citet{Tokunaga2004} superimposed on the CO J=3--2 emission, integrated across all channels showing linear structure (0.67--7.01 km/s). Registration was accomplished by scaling the two images to the same size and then positioning the obvious stellar image of KH 15D on the H$_2$ image at its proper position, as described in the caption to Figure \ref{figure:COmap}. The shape of the CO emission closely resembles the H$_2$ filament to the north, and additionally lines up with the faint H$_2$ emission that re-appears at the southern border of the \citet{Tokunaga2004} image. This gives us confidence that the positional registration is correct and confirms that the star does not lie exactly along the CO outflow axis, although it is very close. 
\label{figure:jet_overlap}}
\end{figure}

By far, the most exciting new result from the radio observations is the discovery in the SMA data of a CO outflow in the J=3--2 transition that matches the near-IR molecular feature so nicely (Figure \ref{figure:jet_overlap}). This CO outflow extends 15$\arcsec$ to both the north and the south, corresponding to a full linear extent of approximately 24,000 AU at the estimated distance of KH 15D. While most CO outflows have wide opening angles, this structure is tightly collimated, with a collimation factor (length:width) of 13.6. Tight collimation is closely correlated with high jet velocities according to \citet{Bachiller1996}. Assuming a relatively modest constant jet speed of 100 km/s, we can estimate the timescale of the jet to be around 550 years. Such a short timescale suggests that we should be able to see changes in the structure of the outflow with repeated observations of the CO J=3--2 line, perhaps over just the six year period between the SMA observations and today. Spatially resolving these structural changes in the future would make it possible to estimate the translational velocity of the CO outflow, which combined with the radial velocities would give the true outflow speed.

The newly detected CO outflow is most certainly associated with the previously known H$_2$ feature. As may be seen in Figure \ref{figure:jet_overlap}, the morphology of the two filaments is precisely the same in almost every detail. The main difference is that the full CO filament can be seen in the radio whereas the southern portion of the near-IR outflow is obscured by an intervening cloud. We have carefully measured the postion angle of the well-delineated outflow and find it to be 17.6 $\pm$ 0.3 degrees. 

Is this filament associated with KH 15D? The only reason to doubt that is the small, but real, offset between the optical/near-IR image of the star and the central line of the outflow, which is easily seen on Figure \ref{figure:jet_overlap} and amounts to about 1$\arcsec$ or 800 AU. By contrast, the following facts argue that the KH 15D system, if not the optically visible binary, is the source of the molecular filament: 1) The star is located almost at the central point along the CO filament, showing that the lack of a ``counterjet" in the near-IR image is an extinction effect. 2) The radial velocity of the feature is consistent within the errors with the systemic velocity of KH 15D and may, in fact, prove to be an important element in ultimately establishing that systemic velocity for the full dynamical model to come. 3) The position angle of the CO outflow, 17.6 $\pm$ 0.3 degrees, agrees remarkably well with the position angle of the polarization of light from KH 15D during eclipse measured by \citet{Agol2004}, 17 $\pm$ 1 degrees. Assuming the scattered light comes from forward scattering off of the CB ring \citep{Silvia} one would expect it to be polarized in a direction perpendicular to the ring/disk, i.e. in the same direction that one would expect an outflow to emerge. 4) The small range in velocities for what must be a high speed outflow, given its tight collimation, indicates that the outflow must be nearly perpendicular to our line of sight, again the orientation one would expect given that the ring/disk must be nearly edge on to cause the optical obscuration effects we observe.

In short, it is hard to believe that the spatial, kinematic, and orientation coincidences between KH 15D and the molecular filament are caused by some chance superposition of two unrelated physical entities. On the other hand, the small offset of the star from the central line of the CO element is definitely real and demands explanation. Close inspection of Figure \ref{figure:jet_overlap} or, more definitively, the original H$_2$ images displayed by \citet{Tokunaga2004} do show a wisp of nebulosity that emerges directly from the location of the binary star in a nearly exactly northward direction, then appears to bend and merge into the main molecular feature with a 17.6 degree position angle. This detailed feature of the outflow may be telling us something about the jet launching mechanism in this source (and in binary T Tauri stars in general). On the other hand, there is no hint of that feature in the radio map, so it could be an artifact. A speculative possibility is that the jet is actually associated with the putative third luminous object in the system mentioned in the Introduction. In a study of spectroscopic binaries with periods between 12--30 days, \citet{Tokovinin2006} found 34\% had a tertiary companion. However we note that this occurrence rate was found to decline with increasing orbital period, and that KH 15D's 48.37 day period is over 50\% longer than the 30 day upper limit for the sample. As discussed above, the time scales on which the outflow features are expected to evolve are sufficiently short that it is probably worth revisiting the source to obtain high spatial resolution modern images of the near-IR and radio molecular filaments. 

\section{Summary}

\begin{figure}[h!]
\epsscale{0.95}
\plotone{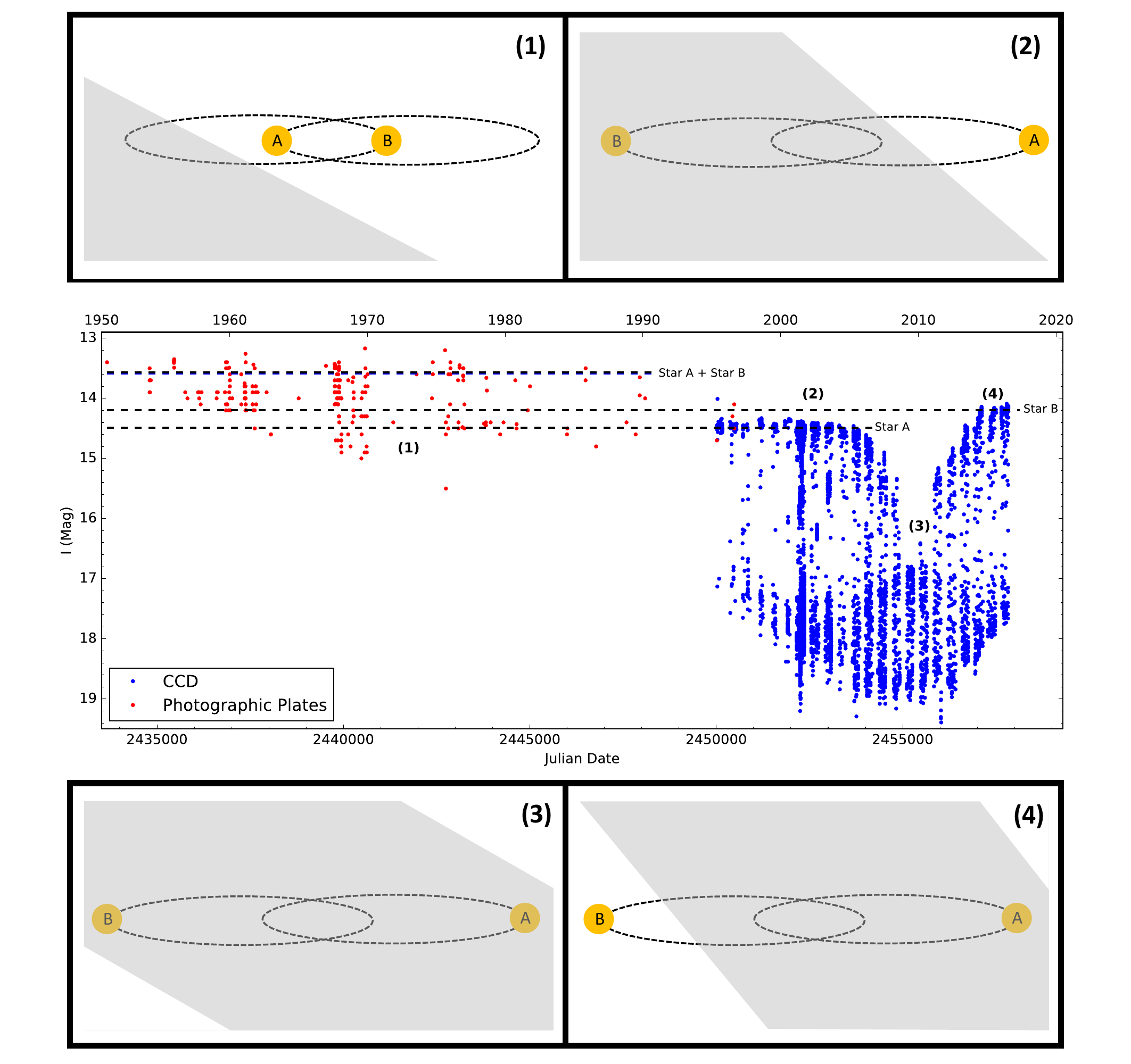}
\caption{The I band light curve of KH 15D, showing how the brightness of the system has evolved over time from 1951 to 2017. The blue data are CCD photometry from Table \ref{table:photometry}, while the red data are photographic photometry from \citet{JohnsonWinn2004, Johnson2005, Maffei2005}. The measured mean brightness of Star A, Star B, and the combined light is shown by the black dotted lines. The sum of the Star A magnitude and Star B magnitude is plotted as a blue dotted line, and overlaps the underlying black line indicating the close agreement between the photographic and CCD data. For each time period (1)--(4), a cartoon of the system in its bright phase is illustrated, demonstrating the changing projection of the disk on the plane of the sky.
\label{figure:intro_figure}}
\end{figure}

Due to the steady, continuing precession of its CB ring, KH 15D has entered a new phase of its light curve evolution as seen from Earth, in which Star B is completely unocculted during each apastron passage (see panel 4 of Figure \ref{figure:intro_figure}). This has allowed us to accurately determine, for the first time, the luminosity and color of each star. From these data, and an assumed reddening and distance to the object, we also derive effective temperatures and radii for the stars as well as model-dependent masses and age. We find that Star B is slightly more luminous and of slightly higher effective temperature than Star A, while having about the same radius. The stellar data are consistent with coeval stars in the 2.5--5.0 Myr range of about 0.7 M$_\odot$ each, with Star B being slighly more massive. When radial velocities of Star B are available it will be possible to expand on the model of \citet{Winn2006} to obtain stellar parameters of much higher accuracy, which should prove extremely useful in testing models of pre-main sequence evolution.

We also report the first sensitive radio observations of the system, done with interferometers (ALMA and SMA) because of the complex surroundings. These data allow us to place a limit on the disk mass of 1.7 M$_\mathrm{jup}$, since we were unable to confirm any thermal emission from the KH 15D disk with the higher sensitivity observations from ALMA. The SMA data has revealed a CO J=3--2 outflow associated with the KH 15D system, which is coincident with the filamentary H$_2$ emission discovered by \citet{Tokunaga2004} that extends northward of the binary. The SMA data also shows the existence of a counterjet extending southward from the source. The central velocity of the jet matches well with the systemic velocity of the binary determined from the optical spectroscopy. Also, the alignment of the jet is in perfect agreement with the position angle of the polarization of the scattered light from the binary. Finally, the small velocity range of the CO jet coupled with its tight collimation indicates that the outflow is nearly perpendicular to our line of sight. In all respects, the CO outflow matches expectation for an outflow emanating from the KH 15D system. Nonetheless, there is a small, but real, offset of the optical/IR image of the binary from the central line of the CO outflow, which raises the possibility that the outflow does not come from the binary itself, but from a putative third object orbiting the binary at $\sim$1000 AU or more. 

Future observations that would be useful in elucidating this situation include radial velocity measurements of Star B to constrain its orbit, which should lead to a very high accuracy set of stellar data for the binary components and a strong test of pre-main sequence models. In addition, it may soon be possible to detect changes in the outflow features, which could yield the outflow velocity, so new, high resoultion IR imagery would be useful. Further study of the system during its faint phase may help resolve the issue of whether there is a third source of light associated with the binary and whether it is positionally coincident with the center of the newly discovered CO outflow. KH 15D remains a unique object because of its precessing CB ring and continues to offer possibilities to enrich our understanding of star, and possibly planet, formation not offered by any other known source.

\acknowledgements
The Submillimeter Array is a joint project between the Smithsonian  Astrophysical Observatory and the Academia Sinica Institute of Astronomy and  Astrophysics and is funded by the Smithsonian Institution and the Academia Sinica. We acknowledge the support of the National Optical Astronomy Observatory through Proposal ID: 2016B-0103 which provided some time on the ANDICAM/CTIO 1.3 m telescope, operated by the SMARTS consortium. This paper makes use of the following ALMA data: ADS/JAO.ALMA\#2015.1.01527.S. ALMA is a partnership of ESO (representing its member states), NSF (USA) and  NINS (Japan), together with NRC (Canada), MOST and ASIAA (Taiwan), and KASI (Republic of Korea), in cooperation with the Republic of Chile. The Joint ALMA Observatory is operated by ESO, AUI/NRAO and NAOJ. The National Radio Astronomy Observatory is a facility of the National Science Foundation operated under cooperative agreement by Associated Universities, Inc. One of us (RA) was also supported by a grant from the NASA CT Space Grant Consortium, for which we are most grateful. We thank E. Agol, C. Hamilton and J. Johnson for helpful discussions regarding this investigation and the source in general.


\clearpage

\begin{deluxetable}{ccccccccccccccc}
\tabletypesize{\scriptsize}
\centering
\tablewidth{0pt}  
\tablecaption{VRIJHK Photometry of KH 15D \label{table:photometry}
}
\tablehead{\colhead{Julian Date} & \colhead{V} & \colhead{$\sigma_V$} & \colhead{R} & \colhead{$\sigma_R$} & \colhead{I} & \colhead{$\sigma_I$} & \colhead{J} & \colhead{$\sigma_J$} & \colhead{H} & \colhead{$\sigma_H$} & \colhead{K} & \colhead{$\sigma_K$} & \colhead{Obs} & \colhead{Ref}}
\startdata
2450017.84 & \nodata & \nodata & \nodata & \nodata & 14.45 & 0.011 & \nodata & \nodata & \nodata &  \nodata & \nodata & \nodata & VVO & H05  \\
2450021.75 & \nodata & \nodata & \nodata & \nodata & 14.52 & 0.024 & \nodata & \nodata & \nodata & \nodata & \nodata & \nodata & VVO & H05 \\
... & ... & ... & ... & ... & ... & ... & ... & ... & ... & ... & ... & ... & ... & ... \\
2457834.54 & 15.47 & 0.019 & 14.74 & 0.019 & 14.16 &  0.019 & 13.25 & 0.041 & 12.54 & 0.036 & 12.52 &   0.095 & CTIO & A17  \\
2457835.54 & 15.45 & 0.019 & 14.70 & 0.019 & 14.13 & 0.019 & 13.19 & 0.042 & 12.46 & 0.037 & 12.41 & 0.095 & CTIO & A17 \\
\enddata
\tablenotetext{a}{An excerpt from Table 1. The full version (available electronically) includes all VRIJHK photometry from 1995--2017, as well as the observatory from which the measurements were taken and the paper from which they were accessed. In the electronic version, nights with no available data are represented by magnitudes and errors of 99.99 and 9.999, respectively. The observatory and reference keys are also included in this version.}
\end{deluxetable}

\clearpage
\begin{deluxetable}{lcc}
\tablewidth{0pt}
\tablecaption{SMA Observational Parameters}
\tablehead{
\colhead{Parameter} & \colhead{2010 Oct 13} & \colhead{2011 Oct 25}}
\startdata
No. Antennas & 7 & 7 \\
$\tau_{\rm 225~GHz}$ & 0.04 & 0.08 \\
Pointing center (J2000) &
\multicolumn{2}{c}{
  R.A. $06^{h}41^{m}10\fs18$, Dec. $+09^{d}28^{m}35\fs5$
  } \\
Min/Max baseline  & 15 to 76 meters & 11 to 76 meters \\
Gain Calibrators  & J0532+075 (0.64 Jy) & J0532+075 (0.72 Jy) \\
                  & J0750+125 (0.57 Jy) & J0750+125 (0.67 Jy) \\
Passband Calibrator & 3C454.3 & Uranus, J0927+390 \\
Flux Calibrator    & Callisto & Uranus \\
Synthesized Beam FWHM\tablenotemark{a}
  & $2\farcs2 \times 1\farcs6$, p.a. $35^{\degr}$
  & $2\farcs1 \times 2\farcs0$, p.a. $-16^{\degr}$ \\
rms noise (continuum image)      &  1.3 mJy & 0.9 mJy \\
rms noise (line images)          & \nodata & 87 mJy \\
\enddata
\tablenotetext{a}{natural weighting of the visibilities} 
\label{tab:obs}
\end{deluxetable}

\clearpage
\begin{deluxetable}{l|ccc|c|ccc}
\centering
\tabletypesize{\scriptsize}
\tablewidth{0pt}  
\tablecaption{Brightness of KH 15D Components \label{table:stellar_mags}}
\tablehead{\multicolumn{1}{l|}{ } &  \multicolumn{3}{c|}{Apparent Mag} & \multicolumn{1}{c|}{Extinction (mag)\tablenotemark{a}} & \multicolumn{3}{c}{Absolute Mag\tablenotemark{b}} \\
\colhead{Band}  &   \colhead{Star A} & \colhead{Star B} & \colhead{Both Stars}      &      & \colhead{Star A} & \colhead{Star B} & \colhead{Both Stars}}
\tablecolumns{8}
\startdata
$V$ & $16.039\pm0.003$ & $15.509\pm0.009$ & \nodata & 0.233 & $6.756\pm0.055$ & $6.226\pm0.056$ & \nodata \\
$R$ & $15.257\pm0.005$ & $14.776\pm0.008$ & \nodata & 0.175 & $5.916\pm0.055$ & $5.436\pm0.055$ & \nodata  \\
$I$ & $14.489\pm0.001$ & $14.198\pm0.009$ & $13.57\pm0.03$\tablenotemark{c} & 0.111 & $5.085\pm0.054$ & $4.794\pm0.055$ & $4.17\pm0.06$ \\
$J$ & $13.504\pm0.017$ & $13.285\pm0.010$ & \nodata & 0.066 & $4.054\pm0.057$ & $3.835\pm0.055$ & \nodata \\
$H$ & $12.825\pm0.015$ & $12.570\pm0.013$ & \nodata & 0.044 & $3.353\pm0.056$ & $3.099\pm0.056$ & \nodata \\
$K$ & $12.541\pm0.014$ & $12.421\pm0.022$ & \nodata & 0.027 & $3.052\pm0.056$ & $2.932\pm0.059$ & \nodata \\
\enddata
\tablenotetext{a}{R $=3.1$, E(B-V) $=0.075$ $\pm$ 0.003, extinction law from \citet{Cardelli1989}}
\tablenotetext{b}{d $=800$ $\pm$ 20 pc}
\tablenotetext{c}{\citet{Winn2006}}
\end{deluxetable}

\clearpage
\begin{deluxetable}{lcc}
\centering
\tabletypesize{\footnotesize}
\tablewidth{0pt}  
\tablecaption{Distance and Reddening of NGC 2264 \label{table:cluster_facts}}
\tablehead{\colhead{Reference} & \colhead{Distance (pc)} & \colhead{E(B-V)}}
\startdata
\citet{Walker1956} & $800$ & $0.082$ \\
\citet{MendozaGomez} & $875$ & \nodata \\
\citet{KwonLee1983} & $700$ & $0.07$ \\
\citet{SagarJoshi} & $800\pm75$ & 0.00 -- 0.12  \\
\citet{Perez1987} & $950\pm75$ & $0.061\pm0.021$ \\
\citet{Feldbrugge} & $700\pm40$ & $0.042\pm0.009$ \\
\citet{Neri1993} & $910\pm50$ & \nodata\\
\citet{Sung1997} & $760\pm85$ & $0.071\pm0.033$ \\
\citet{Park2000} & $760\pm35$ & $0.066\pm0.034$\\
\citet{Baxter2009}\tablenotemark{a} & $913\pm40$ & \nodata\\
\citet{SungBessell2010}\tablenotemark{b} & $815\pm95$ & \nodata\\
\citet{Turner2012} & $777\pm12$ & $0.075\pm0.003$ \\
\citet{Kamezaki2013}\tablenotemark{c} & $738^{+57}_{-50}$ & \nodata \\
\citet{Dzib2014}\tablenotemark{d} & 400 & \nodata  \\
\citet{Gillen2014}\tablenotemark{e} & $756\pm96$ & \nodata \\
\enddata 
\tablenotetext{a}{ $v\sin i$ modeling of NGC 2264 stars. Note that this technique has an additional systematic uncertainty of $\pm110$ pc.}
\tablenotetext{b}{Mean distance derived from SED fitting of cluster members}
\tablenotetext{c}{Parallax of water masers associated with NGC 2264}
\tablenotetext{d}{Combination of the following three methods: 1) Fitting radial velocities and proper motions of sources in both NGC 2264 and (independently) HH 124 IRS, a radio cluster thought to be associated with NGC 2264, to an adopted galactic rotation curve. 2) Re-analyzing parallax measurements taken with Hipparcos of NGC 2264 members. 3) A convergent point analysis of NGC 2264 stars with known proper motions and radial velocities.}
\tablenotetext{e}{Derived from eclipsing binary CoRoT 223992193 in NGC 2264}
\end{deluxetable}

\clearpage
\begin{deluxetable}{lcc}
\centering
\tablewidth{0pt}  
\tablecaption{Derived Stellar Properties of KH 15D Components\tablenotemark{a} \label{table:derived_stellar_props}
}
\tablehead{\colhead{} & \colhead{Star A} & \colhead{Star B}}
\startdata
Spectral Type & K7\tablenotemark{ b} & K1\tablenotemark{ c}  \\
(V-I$_\mathrm{C}$)$_0$ & $1.671\pm0.078$ & $1.432\pm0.078$ \\
T$_\mathrm{eff}$ (K) & $3970\pm40$ & $4140\pm155$ \\
BC$_\mathrm{V}$ & $-1.14\pm0.05$ & $-0.95\pm0.16$ \\
Luminosity ($L_\odot$) & $0.45\pm0.03$ & $0.61\pm0.10$ \\
Radius ($R_\odot$) & $1.41\pm0.05$ & $1.52\pm0.16$ \\
Mass ($M_\odot$) & $0.715\pm0.005$ & 0.74$^{+0.09}_{-0.04}$ \\

\enddata
\tablenotetext{a}{Results for adopted cluster parameters of d = 800 $\pm$ 20 pc, E(B-V) = 0.075 $\pm$ 0.003. Effective temperature and bolometric correction are derived from intrinsic V-I$_\mathrm{C}$ color  \citep{PM13}. Solar parameters of M$_\mathrm{bol,\odot}$ = 4.74 and T$_\mathrm{eff, \odot}$ = 5772.0 were adopted for the luminosity and radius calculations \citep{sunProps}. The model-dependent age is 2.5--5.0 Myr for both stars.}
\tablenotetext{b}{ \citet{Hamilton2001}}
\tablenotetext{c}{ \citet{Capelo2012}}
\end{deluxetable}



\clearpage

\clearpage

\clearpage

\clearpage

\clearpage

\clearpage

\clearpage

\clearpage

\clearpage

\clearpage

\end{document}